# Springtide-induced magnification of Earth mantle's resonance causes tectonics and conceals universality of physics at all scales

Mensur Omerbashich

*Physics Department, Faculty of Science, University of Sarajevo, Zmaja od Bosne 33, Sarajevo, Bosnia*
*Phone +387-63-817-925, Fax +387-33-649-359, E-mail: momerbasic@pmf.unsa.ba; cc: omerbashich@yahoo.com*

**Abstract**

I demonstrate two fundamental contributions. First, the Earth tectonics is generally a consequence of the springtide-induced magnification of mechanical resonance in the Earth's mantle. The same mechanism that causes bridges to collapse under the soldiers step-marching makes also the Earth's lithosphere fail under the springtide-induced magnification of the mantle resonance resulting in strong earthquakes. Secondly, by generalizing the above finding onto any body anywhere in all the Universes and at all times, I find that there is no distinction between physics at intergalactic, Newtonian, quantum, and smaller scales. Thus, the so-called constant of proportionality of physics, $G$, is not a constant but a parameter of a most general form: $G = s \cdot e^2$, nonlinearly varying amongst different scales $s$. Any scale-related variations of physics, erroneously recognized as such by Einstein and Planck, are only apparent and arise as a consequence of the Earth mantle's springtide-induced extreme resonance, which is also critically impeding any terrestrial experiments aimed at estimating the final proportionality $G$. Gravitation is explained if simply regarded mechanical and repulsive.

## 1. Introduction: Gravimetric Terrestrial Spectroscopy

I use superconducting gravimeter (SG) 1Hz observations to show that the Earth is a simple forced mechanical oscillator in which resonant magnification of Earth total-mass (mostly the mantle) oscillation occurs. This means that the Earth tectonics and related phenomena are mechanically caused, and not thermo-nuclear-chemically as previously hypothesized. The lunar synodic semi-monthly forcing drives incessantly the long-periodic (between 12-120 minutes) oscillation magnitudes of the Earth's total mass. In addition, total-mass oscillation magnitudes correlate up to 0.97 with seismic energies on the day of shallow and 3 days before deep earthquakes. The forced-oscillator equations for the Earth mantle's usual viscosity, the Earth grave mode and the lunar synodic half-month (14.7655 days) periods, successfully predict the SG-observed 3-days phase lag in gravity-seismicity correlation. The same equations predict the maximum displacement anywhere on Earth as ~9.8 m corresponding to the average movement during the ~M9.5 (gravest) earthquakes. In the proposed model large earthquakes of unspecified type (mostly strike-slip and thrust are used herein) are entirely predictable. Much as a construction engineer is aware of the infamous examples of structural collapses such as when wind or marching soldiers crashed entire bridges, a seismologist of future will want to observe whether the magnified springtide-induced mantle resonance matches the grave mode of oscillation of specific mass bodies in the crust.

Furthermore, modeling the Earth as a Moon-forced oscillator requires an additional-pull correction to gravitation of up to $1+2.66\cdot10^{-11}$ (5.9 GN), due to resonant magnification of Earth total-mass (mostly the mantle) oscillation. In order to demonstrate that the proposed model applies to the whole Earth, I regard my model as being valid to everything and at all times. By doing so, I find the unifying relationship between physics at the Newtonian and Planck scales thus corroborating the universal generalization of the proposed model, as well as validating that model.

References on "tidal triggering of earthquakes" are numerous [1]. For example, [2,3,4] found a correlation between tidal potential and times of occurrence of earthquakes on the lunar synodic time-scale at ~14.8 days, while [5] asserted azimuth-dependent tidal triggering in regional shallow seismicity. However, many of the tidal triggering claims have been disputed. For instance, [6] dismissed [3], stating that only semidiurnal tide has sufficient power to trigger earthquakes. [7] did not even make reference to fortnightly periods in discussing tidal triggering, while [8] questioned [5] proposing that it be tested "whether it is the oscillatory nature of tidal stress – rather than its small magnitude – that inhibits triggering" [8]. A century of "tidal triggering" reports is summarized in [9]: *"The following periodicities in earthquake occurrence have been proposed at one time or another: 42 min, 1 day, 14.8 days, 29.6 days, 6 months, 1 year, 11 years, and 19 years. (…) Yet a Fourier analysis of earthquake time series fails to detect significant spectral lines corresponding to luni-solar periods."*

Twenty superconducting gravimeters (SG) are in use worldwide [10] for studying Earth tides, Earth rotation, interior, ocean and atmospheric loading, and for verifying the Newtonian constant of gravitation [11]. I used gravity (mass acceleration) data from the Canadian SG at Cantley, Quebec. This stable (as found by [12]) instrument is sensitive to about one part in $10^{12}$ of surface gravity at tidal and normal mode frequencies, so it records antipodean earthquakes as small as M~5.5 [13]. To process SG-gravity, from gapped records of 1Hz output, I constructed a non-equispaced filter with 2-sigma Gaussian, and 8-seconds filtering step as recommended by [10].

The state and perturbations of all the Earth masses, including atmosphere, ocean, land, and interior, are of interest here. Since the mass of gas, water, rock, mantle, and core together account for Earth gravity, hence <u>information equals signal</u>. Such a concept is justified by work of [14] who established the usefulness of geophysical noise in energy (wave) information for studying the Earth's interior. Of all SG locations, the Canadian one was the best for this purpose, as it is antipodal to the seismically most active region on Earth – the Pacific Rim. This ensured that the signal strength was maximal. Thus, the whole Earth is studied by means of raw gravity observations (further: gravity observations) that are neither stripped of tides nor corrected for environmental effects. This can be achieved by looking into Earth gravity spectra from the band of Earth all masses' long eigenperiods 12–120 min, or low eigenfrequencies 12–120 cycles per day (c.p.d.). I term my approach *Gravimetric Terrestrial Spectroscopy* (GTS).

## 2. Methods

Gravity spectra were obtained using least-squares spectral analyses (herein called *Gauss-Vaníček spectral analysis* – GVSA) of over ten billion non-equispaced, Gaussian-filtered SG gravity recordings. GVSA fits, in the least-squares sense, data to trigonometric functions. GVSA was first proposed by [15], was first developed by [16,17], and was simplified by [18] and [19]. Magnitudes of GVSA-derived, variance-spectrum peaks depict "the contribution of a period to the variance of the time-series, of the order of (some) %" [16]. Thus, variance spectra, expressed in var%, or power spectra, expressed in dB [20], can be produced from incomplete numerical records of any length. Thanks to its many important advantages, GVSA was a more suitable technique for GTS than any of the typically used tools such as the Fourier spectral analysis [21,22]. GVSA seemed most apt for GTS primarily due to: (i) ability to handle gaps in data [23], (ii) straightforward significance level regime [24], and (iii) distinctive and virtually unambiguous depiction of background noise levels [13]. These advantages make GVSA a unique field descriptor that can accurately and simultaneously estimate both the structural eigenfrequencies and field relative dynamics. Over the past thirty years, GVSA was applied in astronomy, geophysics, medicine, microbiology, finances, etc. See [13] for GVSA references and a blind performance-test using synthetic data, and [22] for an example of superiority of GVSA over the Fourier spectral analysis.

The 2-sigma Gaussian was selected for filtering the SG gravity data in a non-equispaced fashion, meaning that the sub-step gaps were accounted for by Boolean-weighting each measurement. This filter's response stayed well above 90 var% over the entire band of interest, passing all the systematic contents from 1 min to 10 years. At no point in this research did any of the gravity spectra oscillation magnitudes exceed 1 var%, and most of the time they stayed a hundred times beneath that level – in the order of 0.01 var%. The equispaced Gaussian weighting function $w$, used here for filtering of series of step size $\Delta t$ with $(2N + 1)$ elements, for $n$ observations $l(t)$ at the time instant $t$, and for selected $\sigma = 2$, is [21]:

$$w_i(\Delta t, \sigma) = \frac{1}{2\sigma\sqrt{2\pi}} e^{\frac{(i\Delta t)^2}{2\sigma^2}}, \quad (1)$$
$$\forall i = -N, -N+1, \ldots, 0, 1, \ldots, N$$

and the filter [21]:

$$l_j^*(l, w) = \frac{1}{\sum w} \cdot \sum_{i=0}^{N} l_{j+i}(t) \cdot w_i(\Delta t), \quad (2)$$
$$\forall j = 1, \ldots, n$$

becoming a Boxcar case for $w_i = 1 / (2N + 1)$ $\forall$ $i \in \aleph$. The guiding idea behind the non-equispaced filter was to enable rigorous data processing in which no low-frequency information would be lost due to filtering, unlike in equispaced filters (usually applied for Fourier methods), where variation in the original ratio of populated $p_i = \{l_1, l_2...\}$ versus empty placeholders $q_i = \{\}$ is overlooked. Thus, a distortion of the data by contrivance of invented values that must fall on the integer number of steps takes place in cases when equispaced filters are used. When a portion of the record lacks observations, its average should be re-normalized regardless of the choice of the filtering function, as:

$$l_i^{**} = l_i^* / \sum_{\substack{existing \\ data\ pts.}} l_i^* \;:\; \sum_{\substack{existing \\ data\ pts.}} l_i^{**} = 1, \quad (3)$$

where $l_i^{**}$ are re-normalized filtered values.

Based on Jeffreys's rule of thumb ("*In many earthquakes observations of only the horizontal Earth movements during the passage of shear (S) waves can be used to estimate the order of the total released energy.*"[25: 267]), and the Earth taken as a viscoelastic continuously vibrating stopped up mechanical system, for instance [26], I propose the following

> **Physical hypothesis "A":**
> *The ratio of seismic energy $E_S$ and total kinetic energy $E_K$ on Earth is constant.*

Note that most earthquakes used herein were tectonic thrust and strike-slip.

Seismic energy ($E_S$), as that part of the total kinetic energy transmitted by the lithosphere, is normally found from earthquake magnitudes estimated at seismic observatories worldwide. Since the lithosphere makes merely ~2% of the Earth's volume (for comparison mantle makes ~70%) and ~1% of the Earth's mass [27], the hypothesis "A" generally holds. Seismic energy expressed in units of ergs is computed using modified Richter-Gutenberg formula, found in, for instance, [9]:

$$\log_{10} E_S = 1.5 \cdot M_S + 11.4. \quad (4)$$

Physically, magnitudes of gravity field oscillations are proportional to the kinetic energy ($E_K$) needed to displace the Earth's inner masses as the medium, from the state of rest to that of unrest [28]. Therefore, I computed as a relative measure of the Earth kinetic energy the series of simple-average magnitudes ${}^T\bar{\mu}_\omega$ of Earth oscillation at particular frequency $\omega$, as determined from a gravity record spanning a specific period of time $T$ (Week, Day, Hour, etc.). Then for a normal mode of Earth oscillation, the amplitude ${}^T\bar{\mu}_\omega$ of a gravity GVSA spectrum $\mathbf{s}^{GVSA}(\omega)$ is computed as the average of three (minimum spectrum size in GVSA), as:

$$^T\bar{\mu}_\omega = \frac{1}{3}\sum_{i=0}^{2}s(\omega_i),\qquad(5)$$

$$\omega_i = \begin{cases} 1/[(\text{mode period})^{[\text{sec}]} -1^{[\text{sec}]}], & \text{for } i=0 \\ 1/[(\text{mode period})^{[\text{sec}]}], & \text{for } i=1 \\ 1/[(\text{mode period})^{[\text{sec}]} +1^{[\text{sec}]}], & \text{for } i=2 \end{cases}$$

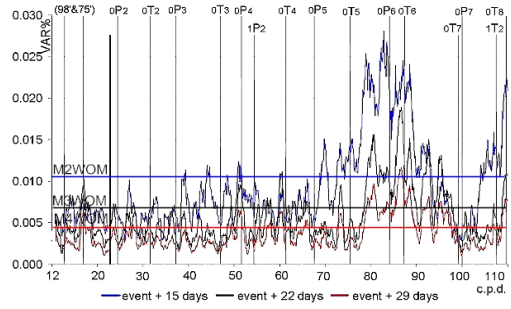

Fig. 1. Gravity variance-spectra as a relative measure of kinetic energy. Mean-weekly oscillation magnitudes $^W\bar{\mu}_E$ (MWOM) from variance spectra of detided gravity at Cantley from 2 (blue), 3 (black) and 4 (red) weeks of data past the M8.8 Balleny Islands earthquake [64], of 25 March 1998, Harvard CMT-032598B. Focus was in Tasman fracture zone between Southeast Indian and Pacific-Antarctic ridges, 700 km east of South magnetic pole along winter ice boundary and 600 km north of George V Coast. East Antarctica is stable Precambrian shield composed of 3+ billion years old metamorphic rocks that did not undergo major change in recent geological times. Normal periods Zharkov model. Filter step 32 s. Resolution 1000 pt.

As a test, Fig. 1 depicts the average oscillation magnitudes $^T\bar{\mu}_E$ (colored lines) over all (here 1,000) spectral points in the band of interest $\mathbf{E}_l \in 12–120$ c.p.d. from 2, 3, and 4 weeks of gravity data past a great earthquake. Earth ringing is thus observed and measured relatively by means of SG gravity variance-spectra for at least six weeks past a >M7.5 event (Fig. 1). This is in agreement with the solid-tide general dissipation rate of 83±45GW established by [29].

### 3. Gravity-seismicity correlation

Based on the starting physical hypothesis "A", linear correlation is sought between the series $\mathbf{X}_{\bar{\mu}}$ of last decade's daily oscillation magnitudes, and the series $\mathbf{Y}_S$ of seismic energies (and for check of seismic magnitudes) from 381 medium-to-large global earthquakes found (by visual inspection) to have excited the Canadian SG record spanning the 1990-ies. Cross-correlation functional values were computed between $n_\omega$ components of vector $\mathbf{X}_{\bar{\mu}}$, and the vector of earthquake energies, $\mathbf{Y}_S$, as [30]:

$$\Omega(X,Y,u) = \frac{\sum_{j=0}^{30}\left\{[X(\tau_0+i\Delta\tau)-\bar{X}][Y(\tau_0+(i+j)\Delta\tau)-\bar{Y}]\right\}}{\sqrt{\sum_{j=0}^{30}[X(\tau_0+i\Delta\tau)-\bar{X}]^2 \sum_{j=0}^{30}[Y(\tau_0+(i+j)\Delta\tau)-\bar{Y}]^2}},\ -1\le\Omega_{x,y}\le 1,\quad(6)$$

where $i = 1, 2, \ldots n_\omega$ ; $u = j\cdot\Delta\tau \wedge j\in \aleph$, $\Delta\tau = 1$ day. Here $\Omega_{x,y}(u)$ is the value of the correlation function $\Omega$ between the components of $\mathbf{X}_{\bar{\mu}}$ and $\mathbf{Y}_S$ lagged by $u$ days (here from $|u| = 0, 1, \ldots 30$ days).

Seismically induced high-frequency Earth oscillation radiates energy in the order of megawatts, *versus* low-frequency (solid-Earth and oceanic) tidal dissipation that amounts to about 2TW, as asserted by [31]. Also, the Earth's incessant high-frequency oscillation, demonstrated by [26] to beat in high eigenfrequencies down to 2.2 mHz or at short eigenperiods up to ~7½ minutes, radiate each day an amount of energy normally released by a single M5.75 - M6 earthquake, as suggested by [32]. For these reasons, magnitudes of Earth high frequency oscillation are not of interest here. Besides, if high correlations could be obtained without using magnitudes of high-frequency Earth oscillation, then this set-up will physically satisfy for the entire natural band as well. Note that [33] and [34] first offered an explanation for the incessant short-periodic Earth oscillation, while [32] offered a more comprehensive explanation of that phenomenon.

The choice of surface magnitudes in Eq. (4) rather than moment magnitudes was not just preferential. Namely, more than 96% of all M6.3+ earthquakes that affected the Cantley SG record were weaker than M7.5 (when the scales based on surface magnitudes start saturating). Hence using surface magnitudes rather than moment magnitudes represents a more stringent approach given that the linear correlation Eq. (6) can be sensitive to a small number of relatively large input values. If high correlations could be obtained using surface magnitudes, then this set-up will satisfy for moment magnitudes as well.

Pairs ($X_{\bar{\mu}}$, $Y_S$) of physically dependent values are not random samples from a bivariate normal distribution as given by [35], so confidence intervals for correlation coefficients cannot be computed [36]. $X_{\bar{\mu}}$ values obtained always from a large sample of up to 86,400 normally distributed gravity measurements per day are not normally distributed either, instead those values follow $\beta$–distribution, as established by [37]. In addition, seismic magnitudes used in the vector $Y_S$ follow the Boltzmann distribution, as reported by [38]. Hence, no statistical tests of Eq. (6) exist.

The only meaningful test is the physical requirement that the value of the correlation function be the largest for lag equal to zero. Therefore, cross-correlations of Eq. (6) were computed between $X_{\bar{\mu}}$ along normal mode periods, and $Y_S$. For this I used three geophysical Earth models: Jeffreys-Bullen "B", a 1967 model based on the compressibility-pressure hypothesis [25], Dziewonski-Gilbert UTD124A', a 1972 model containing sharp density discontinuities [39,40], and Zharkov's 1990 model [41]. I selected M6.3 as the optimal cut-off magnitude so as to avoid interference due to weaker earthquakes [26], and to reduce the computational load. The cut-off magnitude was lowered to M6.0 for deep (> 400 km) earthquakes so as to increase the sample size to over 50.

More than 15,000 computations of Eq. (6), between diurnal oscillation magnitudes $^D\bar{\mu}_\omega$ (Eq. 5), obtained from over 200 million gravity measurements, and global >M6.3 seismicity have returned high correlation values. For the three Earth-models respectively (chronologically) the correlation values reached 0.45, 0.39, 0.39 on the day of the shallow earthquake (Fig. 2(*a*)) and 0.63, 0.65 and 0.67 at three days before the deep earthquake (Fig. 2(*b*)), as well as 0.97 in case of deep earthquakes (likely occurring along the astenosphere-mesosphere interface) at 300-400 km depths, for all three models.

In all cases (models), the strongest response was in mantle-sensitive $_0P_7$ and $_0T_7$ modes, while the lithosphere-sensitive $_0P_2$ and $_0T_2$ modes also returned high correlation values with deep earthquakes (Fig. 2(*b*)). Depth-separation revealed a three-days delay in correlation (Fig. 2(*b*)). In a statistical study of 7,359 deep ≥M3 earthquakes, [42] speculated that a delay in deep rupture might be in effect, without proposing the duration of such a delay. Then given $^D\bar{\mu}_\omega$ are diurnal averages, if it exists, I assume that for most earthquakes such a deep-rupture delay is either $\Delta t_1 <<$ 1/2 day, or $\Delta t_2 > 1$ day. As deduced from the variance spectra (but not the power spectra) of SG gravity, the gravity-seismicity correlation has an absolute maximum for periods ~821 s or eigenfrequencies ~1.22 mHz, as found by [13].

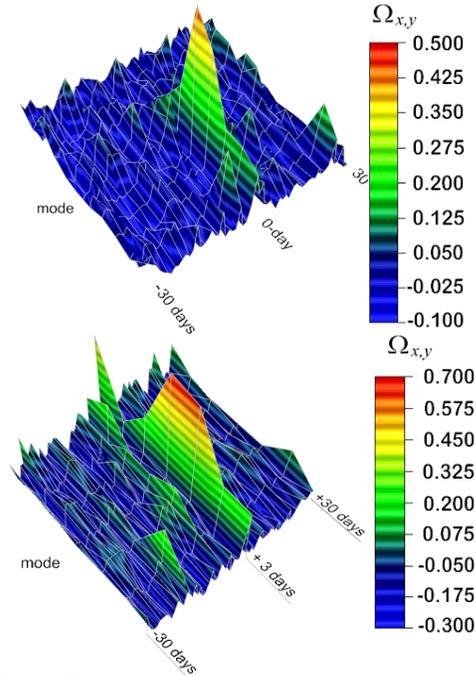

Fig. 2. Top panel (*a*): surface of change of correlation values $\Omega_{x,y}$ between 1991-2001 $^D\bar{\mu}_\omega$ for poloidal periods of Jeffreys-Bullen "B" model, and seismic energies from ≥M6.4 earthquakes. Normal mode periods ordered monotonically from longest (farther) to shortest (closer) long eigenperiods. Bottom panel (*b*): time-period-correlation surface plot, deep seismicity. Surface of change of $\Omega_{x,y}$ between $^D\bar{\mu}_\omega$ and 54, 1991–2001 deep earthquakes M6+ in seismic energies, from variance spectra of SG gravity. Periods 810–833 s, 1 s increment. Time-scale lag in days.

In all three Earth-models tested, gravity-seismicity correlations were higher when seismic energies were used rather than seismic magnitudes, as well as when variance spectra were used rather than power spectra. Also, the more recent model produced higher correlation in case of deep earthquakes. Superconducting gravimeters can sense globally minute (in the order of 0.01 var%) mass re-distributions of the inner masses, and contrary to some opinions, such as for instance [43]), depending on the choice of signal and noise the removal of atmospheric and other environmental effects is not necessary for all purposes. Note that, while the gravity-seismicity correlation Eq. (6) is frequency-dependent, [44] showed that the relationship of local pressure (here the second-largest information constituent) and SG-sampled gravity is independent of frequency and epoch. In addition, it can be seen from [45] that the air pressure- and gravity-spectra are not expected to show any resemblance. Finally, the amplitude of the theoretical tidal gravity signal, observed here at 0.068 c.p.d. reaches at least 1–2 µGal according to [46], corresponding to a change of 5–10 mm vertically or 1.2 –2.5 metres in latitude. This is well within the SG accuracy as given in [12], so that any signal possibly found in the spectra of SG data cannot be exclusively due to geodetic effects of either height or latitude.

## 4. Spectrum of total-Earth long-periodic oscillation

I next spectrally analysed all the decadal $^D\overline{\mu}_\omega$ time-series computed along each long normal mode of the Zharkov model [41] as the most recent of the three Earth models used.

Correlations (Eq. 6) were checked by methodology, i.e., by using: (i) three geophysical Earth models, (ii) surface instead of moment seismic magnitudes, (iii) the lowermost part of the Earth natural band, (iv) variance- *versus* power-spectra, and (v) seismic energies *versus* seismic magnitudes. In order to check the extracted periodicity of the *spectrum of the spectra* of gravity, I look at the entire computational procedure as a filter, and compute its magnitude-frequency response. Of concern here is the fact that filters can enhance or reduce spectral amplitudes of certain frequencies.

Response in var% of the processing viewed as a filter, i.e., of a data processing procedure based on spectral analyses should determine whether any classical noise, naturally measured by variance, was dominating the extraction of the springtide periodicity from gravity spectra. To compute this response, white noise was fed to the computational procedure. I thus processed a "month long" test-record containing random numbers between (0, 1), limits inclusive. Figure 3 shows that the processing acted as a low-pass filter in the 12 days – 200 days interval. Hence, ~15 days periods most likely are not a consequence of filter amplification.

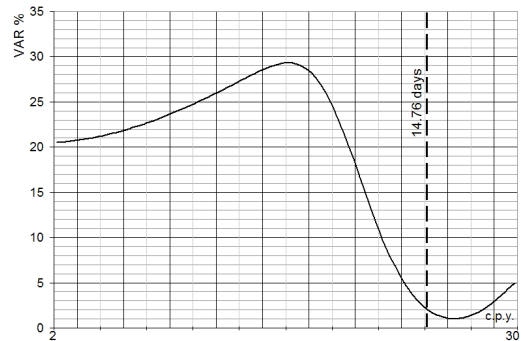

Fig. 3. Magnitude-frequency response of the processing viewed as a filter.

After enforcing the theoretical solar semi-annual period $S_{sa}$= 182.6211 days, all 16 normal $^D\overline{\mu}_\omega$ series are periodic with 14.71 days (Fig. 4). This period is in agreement with "tidal triggering" reports of 14.8 days; see Introduction.

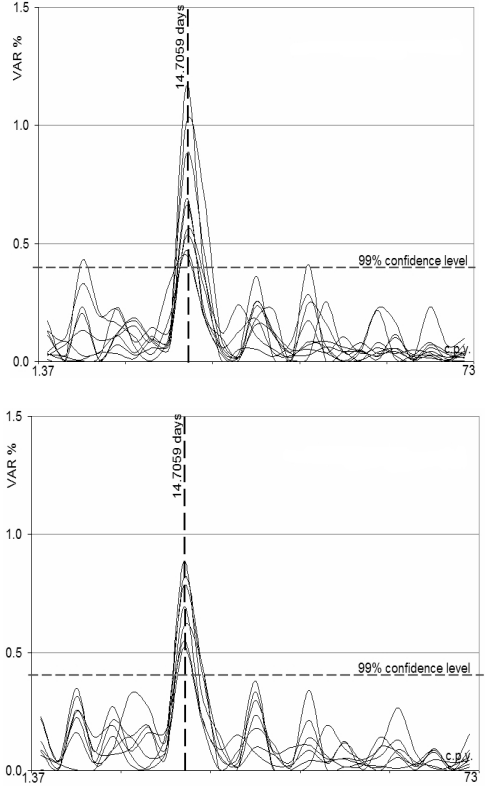

Fig. 4. Lunar synodic periodicity of Earth's daily oscillation magnitudes. Along all poloidal (top panel) and torsional (bottom panel) normal mode periods, Zharkov model. Enforced theoretical solar semi-annual $S_{sa}$ = 182.6211 days.

The maximum magnitudes on Fig. 4 are limited by the frequency-magnitude response of the processing designed to suppress the impact of classical noise on spectral magnitudes at long periods of around half month (Fig. 3). Although largely limited by the processing, the 14.71-day period still exceeds the 99% confidence level (Fig. 4), indicating a high strength of this period. All other periods shorter than 13 days and longer than 16 days, seen on Fig. 4 as not reaching the 99% confidence level, are artifacts of the processing.

If the springtide causes the 14.71-day period, its estimate should get fully enhanced in a series of complete-Earth's $^D\overline{\mu}_E$ obtained over all spectral lines in the band of interest $\mathbf{E}_l \in$ 12–120 c.p.d. This should improve this period's estimate to about the size of the filtering step, since $^D\overline{\mu}_E$ reflects the dynamics of the total Earth (mostly the mantle) as affected by the springtide. In that case, it should also be expected for the accuracy of the estimate of the lunar synodic period to be improved even further – by simply increasing the spectral resolution.

A *lunar synodic month* of ~29.5 days is the mean interval between conjunctions of the Moon and the Sun [3]. It corresponds to one cycle of lunar phases. A more exact, empirical expression that is valid near the present epoch for one lunar synodic month is based upon the lunar theory of [47]:

$$\left. \begin{aligned} T_{syn}\,[\text{m.s.d.}] &= 29.5305888531 + \\ &+ 2.1621 \cdot 10^{-7} \cdot T_{t.d.t.} - \\ &- 3.64 \cdot 10^{-10}\, T^2_{t.d.t.}\,; \\ T_{t.d.t.}\,[\text{JC}] &= (\text{JD}\ 2{,}451{,}545)/36{,}525\,, \end{aligned} \right\} \quad (7)$$

where $T_{syn}$ is measured in *mean solar days* (m.s.d.), and $T_{t.d.t.}$ in Julian days (JD) and Julian centuries (JC) of *Terrestrial Dynamical Time* (T.D.T.) that is independent of the variable rotation of the Earth. Any particular phase cycle may vary from the mean by up to seven hours [47]. It thus sufficed for all purposes to compute spectra with a resolution better than 3½ hours locally around the period of interest. Using the 2,000 pt spectral resolution in this case enables to claim spectral estimates to better than 1.3 h.

The $^D\overline{\mu}_E$ series turned out to be periodic with 14.77876 and 182.61419 days (Fig. 5). The first period is longer than the lunar astronomical fortnightly $M_f$ = 13.66079 m.s.d. by 26h 49m 53s, and longer than the closest theoretical fortnightly of 14.76530 days by 19m 23s. Note that the latter fortnightly tide is ten times smaller in amplitude than the strongest tide $M_f$ excited by the largest tidal potential $P_2^0$, as given in [46]. The second period extracted is the astronomical semi-annual solar, to within 9m 57s. After the removal of the theoretical solar period $S_{sa}$ = 182.6211 days by using GVSA, the lunar period's estimate is 14.76053 days, at a high 26.1 var% and over 95% confidence level.

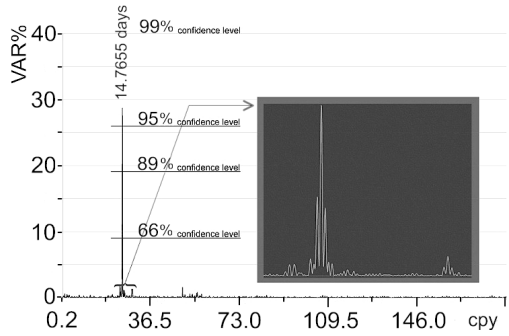

Fig. 5. Lunar synodic half-month periodicity of Earth total mass oscillation. Shown is variance-spectrum of Earth total-mass mean-diurnal magnitudes, obtained from variance-spectra of SG-gravity after the removal of theoretical solar semi-annual period. Data span 10.3 yr. Band 2 day – 5 yr. Spectral resolution 50,000 pt.

Increasing the spectral resolution to 50,000 pt yielded an estimate to better than 3 minutes of the above period, as 14.7655 days at 95% confidence level. According to Eq. (7), this represents an accuracy of ~17 seconds or twice the size of the filtering step. Thus, a considerably improved estimate of the periodicity of the Earth gravity oscillation can be obtained when complete environment information is used, as well as after the spectral resolution is enhanced; 25 times in this case.

The lunar synodic semi-monthly and solar semi-annual periods, as the only periods from the one-day to ten-year period interval cannot represent a residual from data processing as no systematic noise was cleared from the record to begin with. Also, the processing did not amplify these two periods as it, in fact, was designed to suppress all the contents at around half a month (Fig. 3). Thus, the two periods are not in the SG data per sé, and the entire data, i.e., the field that they sample, oscillate with those two periods. This sort of sensitivity was attainable due to SG accuracy (as discussed earlier) appreciably exceeding the ratio between the force of Earth gravity and the maximum lunar tidal force, of ~$1.14:10^{-7}$ [48]. Obviously, since we speak of gravity, the inclusion of gravity components that are due to non-solid Earth masses such as the atmosphere, as well as of those in the deep interior had little importance in inflating these two periods – due to low density and due to distance from the lithosphere, respectively. Here, normal eigenperiods were used because they are the natural beat periods of all Earth masses (further: *Earth total mass*, mostly the mantle) that comprise gravity and cause Earth oscillations in general, not just Earth free oscillation. I therefore checked the periodicity obtained along all normal mode periods (Fig. 4), against the result from using complete environment information (Fig. 5). As seen above, this yielded the lunar period with an astounding accuracy, confirming in effect the starting premise on both the SG accuracy and GTS validity.

## 5. Earth as a Moon-forced mechanical oscillator

Based on SG observations, I define the following

> **Model:**
> *Earth is a viscous, stopped up mechanical oscillator, forced externally mostly by the Moon's orbital period due to which the states of a maximum mass (gravity) oscillation magnification and a maximum stored potential energy occur.*

On a Moon-forced Earth, the planet's total mass (mostly the mantle) oscillates with springtide periods, so Earth oscillation is not just free but constrained as well; meaning rather than observing the response of the Earth as a system under exclusively free (particle) motion, as it has been done in geophysics so far [41], this specific model requires that the response of the Earth system under forced motion be observed instead. Then in such a global forced oscillator, first proposed by [49], where the damping force is proportional to the velocity of the body [28], Earth grave mode $_0P_2$ is the system normal period, and the lunar synodic period is the system forcing period due to which the states of a maximum oscillation magnification and a maximum stored potential energy occur [50].

Let us substitute in the mechanical oscillator equations which can be found in [50] the $T_o$ = 3233 sec (Zharkov model) grave mode as that which makes the normal $\omega_o = 2\pi / T_o$, and SG-measured $T_{max}$ = 14.7655 days as that which makes the maximum magnification forcing frequency $\omega_{max} = 2\pi / T_{max}$. Therefore, the spectral spread of the system response about the lunar resonant frequency $\omega = \omega_{max}$ is then obtained for the characteristic mantle viscosity given in [51], of ~$10^{21}$ Pa s, as [50]:

$$Q = \frac{1}{\sqrt{2 - 2(\omega/\omega_o)^2}} = 0.70711 \cong \frac{1}{\sqrt{2}} \quad (8)$$

and the phase shift of the (response function of the) steady state solution of the Earth displacement as [50]:

$$\phi(\omega_{max}) = \arctan\left(\frac{1/Q}{\omega_o/\omega - \omega/\omega_o}\right) = 0.205 \quad (9)$$

of the forcing period $\phi(\omega_{max})$ = 14.7655 days, or 3.03 days. This phase is the time by which the displacement lags behind the driving force [50]. The theoretical value Eq. (9) of the model-Earth agrees well with the observed 3-day delay in the here discovered gravity-deep seismicity correlation for the real Earth (Fig. 2(b)).

Let us now use the same periods as in the above to obtain the maximum particle displacement on or within the Earth due to magnification of forced oscillation of total Earth mass (where mantle makes ~70% of volume), i.e., of the Earth gravity, as [28,50]:

$$X(\omega) = \frac{F_o}{k} \cdot \left(\frac{\omega_o/\omega}{\sqrt{(\omega_o/\omega - \omega/\omega_o)^2 + 1/Q^2}}\right), \quad (10)$$

where the Earth-Moon maximum gravity force $F_{perigee}$ = 2.2194·$10^{20}$ N is the maximum forcing amplitude $F_o$, and $k = m_E \cdot (\omega_o)^2$ is the system spring constant. For the Earth mass $m_E$ = 5.9736·$10^{24}$ kg (cf. NASA), and $_oP_2$ values of three Earth-models respectively: 3223 s, 3228 s, 3233 s, I obtain the maximal displacement Eqs. (8) - (10), as [50]:

$$x(t) = X(\omega_{max}) \cdot \cos[\omega t - \phi(\omega)]. \quad (11)$$

For shallow earthquakes, where $t = t_o = 0$, $\phi(\omega) = 0$, I obtain 9.78 m, 9.81 m, 9.84 m, respectively (chronologically). For deep (<400 km) earthquakes, where $t = t_o = 0$, $\phi(\omega) = \phi(\omega_{max})$, I obtain 9.57 m, 9.60 m, 9.63 m, respectively (chronologically). Estimates of the grave mode $_oP_2$ from the three Earth-models are seen as increasing by ~5 s per decade, probably due to the Moon receding from Earth (tidal friction), and to totality of all other minor factors such as accretion of cosmic particles and so on. Consequently, it appears that the maximum mass displacement on the herein proposed model-Earth increases steadily at a rate of ~3 cm/decade. This substantiates physically the correlation between the Earth gravity and deep earthquakes, here observed as ever improving decade to decade. In case of shallow earthquakes however, the here-observed correlation trend is seen not as clearly. On the real Earth, this could be due to the lithosphere's rigid-elastic environment being subject to stress-strain build-up and thus the apparent tidal inhibiting. This would then also be in agreement with [52], and also with a general understanding that the size of large strike-slip earthquakes (in this study the most-used type) is not related simply just to the amount of stress-strain buildup [53]. In addition, [54] demonstrated recently that the earthquake triggering is a phenomenon not exclusively due to the release of static strain by foregoing earthquake; instead, seismicity gets activated also by direct shaking, i.e., mechanically. For one to be able to explain this too the Earth must be regarded as a closed mechanical oscillator.

Any stress-strain build-up is virtually absent in the mantle where therefore a forced-oscillator model seems to be able to satisfy the observed trending of the gravity-deep seismicity correlation. Note that during the great ~M9.3 Sumatra earthquake of December 2004, numerous reports have put this displacement at ~10 m average. Then ~M9.5 seems to be the strongest earthquakes possible on an Earth modeled as a mechanical oscillator made (by the springtide-induced resonance) to vibrate to a critical magnification resulting in a structural collapse i.e. an earthquake.

The springtide exerts pull on all Earth masses. In case of the mantle's plastic environment, this would affect the mantle where the mantle is weakest, i.e., along the 300-400 km deep astenosphere-mesosphere interface. The absolutely highest correlation was at the ~821 sec eigenperiod in case of the Pacific Rim. Inserting $T_0$ = 821 sec into Eqs. (8) and (9) yields that region's own ~18.5 h phase shift. This is a real-Earth constraint on the proposed model, as the < 1-day phase shift, if applicable to real Earth, would contribute to the facts that (*a*) shallow earthquakes correlated with Earth oscillation magnitudes on the day of the earthquake and (*b*) this correlation was worse than in the case of deep earthquakes.

According to the Moon-forced Earth oscillator model, when a tectonic structure (mostly the mantle and to a lesser degree the crust) starts resonating in harmony with the Moon-Earth orbital tone, the material of which those structures are composed crumbles, resulting in an earthquake. The forced oscillator model also agrees with the well known fact that shallow and deep earthquakes are observed as belonging to two different processes: plastic and rock environments have radically different structure and therefore considerably different structural eigenfrequencies as well – the longer the object's eigenperiods the shorter the time for the object under forced oscillation magnification to fall apart. Then the forced global oscillator model could perhaps apply to deep seismicity.

As [55] recently found, mantle melts take only a few decades to generate, transfer, accumulate and erupt, opposite to previous estimates of ~$10^3$ yrs. The herein proposed model could perhaps offer a mechanism for rapid transport of mantle melt: in the proposed model tectonics is merely a consequence of the resonating mantle's fast dynamics at a swift rate that equals the time which takes for the (3-days) phase to start affecting the SG sensitivity.

In the realm of the proposed simple model, I thus demonstrated the following

> **Rule:**
> The ratio of seismic energy $E_S$ and total kinetic energy $E_K$ on Earth is constant.

Based on the above-defined model and its *own rule* (a claim that holds only under some strict conditions; a law in a non-strict sense), I propose the following

> **Physical hypothesis "B":**
> Earth tectonics and inner-core differential rotation arise due to respective upward and downward continuations of the springtide-induced magnification of the resonant dynamics of the mantle.

I leave the physical hypothesis "B" untested, for testing it would require that each tectonic structure's grave mode, phase shift, and structural eigenfrequencies are all known, which puts such testing well beyond the scope of this paper.

Importantly, Eqs. (8) - (11) produce nonsensical results for the Earth–Sun system. That is in agreement with the above forced global oscillator model, since, unlike the Moon the Sun "orbits" about the Earth only apparently. This further legitimizes the proposed model by providing relevance for the real Earth.

## 6. Discussion of the model

As the Earth masses bulge during the springtide, their oscillatory movement attains its local maximum amplitude during each new and full Moon. As seen above, I detected these conjunction and opposition peaks in the spectrum of the spectra of SG gravity. No intermediary that would either introduce or amplify the two periodicities could exist, since the perfect periodicity of the $^D\overline{\mu}_E$ series is naturally superimposed onto the periodicity of any information constituent alone and hence of the $^D\overline{\mu}_\omega$ series too. This voids the possibility for any intermediary to introduce the two periods anew. This causality, along with the above theoretical delay, which I also observed experimentally as shown above, and the maximal well-known displacement, which I also obtained theoretically as shown above, constitute a *sufficient condition*, while the $^D\bar{\mu}_\omega$ – seismicity correlation at mantle- and lithosphere-sensitive modes constitutes a *necessary condition* for the real-Earth's lithosphere and mantle to respond to the two periods by rupturing. Rather than, as speculated by some, directly and simply "triggering" the earthquakes, the revealed periodicities in the proposed model either trigger the Earth oscillation itself, or add to the oscillations that were already triggered by earthquakes. Then, the mechanism behind the discovered correlation can be either (*i*) the direct excitation, or (*ii*) an excitation through earthquakes that have been, in turn, triggered by long periodic tides. Based on the computations of Eqs. (8) - (11), and on the ideas of [6] and [7] that are here used as the proposed model's refinements, I discard the latter option (*ii*) above. Then the proposed model does not allow for the so-called "tidal triggering of earthquakes" to exist as such. Furthermore, in reality, oscillatory nature of fault rupturing has long been established, for instance [56], as well as certain related phenomena apart from tides, which can trigger earthquakes, for instance [57].

Thus, the Earth modeled as a forced mechanical oscillator could explain the Earth tectonics. This would require that shallow earthquakes do not depend exclusively on stress-strain conditions, and it would allow for actual prediction of medium-to-large earthquakes. Such prediction would need to determine structural eigenfrequencies of the separate (tectonic) mass bodies and systems, establish which faulting types are most sensitive to the seismicity–gravity correlation (Fig. 2), and subsequently monitor under springtide orbital regimes the structural eigenfrequencies of tectonic masses of interest. The basic idea behind such a concept of earthquake prediction is the simplest one of all, and it resembles the infamous example of soldiers marching across a bridge that subsequently fails.

Given the observed perfect lunar synodic periodicity of the complete gravity field, the Moon forces the total mass (mostly the mantle) of the herein proposed model-Earth. This is also in agreement with numerous reports from various disciplines where the lunar synodic period alone was reported in superimposed-epoch analyses of natural datasets, see, for example [58,59].

According to the forced global oscillator model, all gravitational considerations performed in the vicinity of the Earth, such as those aimed at determining $G$ [11], had failed to account for lunar magnification of total-mass (gravity) oscillation during the considerations, Eq. (7), and at the concerned location, $\omega_o / \omega_{max}$, as well as failed to take into account the (varying) $_oP_2$ of the Earth. The maximum magnification is [50]:

$$M(\omega) = \frac{2Q^2}{\sqrt{4Q^2 - 1}}, \qquad (12)$$

which for the maximum values gives a scaling factor of $M(\omega_{max}) = 1 + 2.66 \cdot 10^{-11}$. The herein used total-mass representation of the Earth results in measuring the natural period of the Earth total mass as $T_o' = T_o + \varepsilon_\omega = 3445$ s $\pm 0.35\%$ where uncertainty is based on 1000 pt spectral resolution. This estimate is in agreement with [60]; see Appendix.

Equations (7) - (12) can be used to scale for magnification all Earth-mass considerations. The $2.66 \cdot 10^{-11}$ scaling factor translates into the maximum force at perigee, of $5.9 \cdot 10^9$ N. Then the mantle's springtide-induced resonance magnification could be responsible also for the unmodeled portion of Earth's nutation, of 10-50 milli-arc-second and presumed to be resonant-periodic with ~30 days period [61]. Finally, if the real Earth could be represented to a good approximation (say, such which would make all the terrestrial determinations of $G$ consistent) by a Moon-forced oscillator, the Earth would then necessarily owe (a part of) its magnetic field to the forcing oscillation [50].

In the realm of astronomically forced oscillators (intergalactic scales), rescaling of gravitational force might be deemed appropriate for all mass considerations within each tidally locked system of two or more heavenly bodies.

## 7. Corollary

I deduce that the lunar forcing Fig. 5 enables the natural vibration of the Earth total-mass (mostly the mantle). Then in the limiting case of Eq. (12), when $T_{max}$ = 14.7655 days period is used along with the natural period of the Earth total mass, an over-scaling (numerical) relationship is asserted between physics at the Newtonian *vs.* Planck scales, i.e., between the Newtonian constant of gravitation (or: constant of proportionality of physics at Newtonian scales), $|G_{Newton}|$ = (6.6742 ± 0.001)·10⁻¹¹, c.f. NIST.GOV, and Newtonian constant of gravitation over $\hbar c$ (or: constant of proportionality of physics at Planck scales), $|G_{Planck}|$ = (6.7087 ± 0.001)·10⁻³⁹, c.f. NIST.GOV, as:

$$G_{Newton} \pm \varepsilon_1 \to G_{Planck} \cdot c^3 \cdot \frac{\omega_o}{\omega_{max}}\bigg|_{Earth-Moon} \quad (13)$$

$$\varepsilon_1 \in \mathfrak{R}$$

suggesting that there is no real distinction between physics at the Newtonian and Planck scales. If the relationship (13) is real, its analytical uncertainty $\varepsilon_1$ must depend only on the uncertainties of $G$ and $\omega_o$.

Analogously, there is no need for invoking the Riemannian geometry in a global forced oscillator. Within the scope of the proposed model, the existence of new frequency-space approaches to the unified theory could be feasible *via* using non-relativistic quantum mechanics alone – by equating the concepts of Earth mass and charge whereas the charge is subject to the lunar forcing just like all the masses that comprise the Earth charge are, Fig. 5. [49] maintained that reality is completely describable within frequency-space alone, and that permanent vibrations (revolutions) of all matter regardless of scale were linearly misconstrued to mean permanent free falling of all matter. In a sense, [49] was an original proponent of the string theory. Newton attached units to the constant of proportionality $G$ in order to complete his theory. As viewed in the here-proposed realm, that close occurs in order for the Newton's (Einstein's) linear misconstruction of the frequency-space reality to be satisfied. Therefore, regarded as a ratio, $G$ is a quantity that carries no units. In the same sense, $c$ represents a unitless ratio, as well. Thus, the proposed model is more general than the Einstein's so-called "general theory of relativity".

Given the measured value $\omega_o$', the following numerical relationships naturally hold everywhere in the vicinity of Earth; like in (13), units are obviously of no concern:

$$G_{Newton} \cdot c \cdot \frac{\omega_o}{\omega_{max}}\bigg|_{Earth-Moon} \to e^2 \quad (14)$$

with ±0.14% analytical uncertainty, and

$$G_{Planck} \cdot c \cdot \frac{\omega_o}{\omega_{max}}\bigg|_{Earth-Moon} \cdot 10^{28} \to e^2 \quad (15)$$

with ±0.40% analytical uncertainty. From expressions (13) - (15) follows the refinement of the resonance ratio of the Moon-forced Earth's total mass, as:

$$\frac{\omega_o}{\omega_{max}}\bigg|_{Earth-Moon} \cdot 10^{-28} \to c^{-3} \quad (16)$$

with ±0.07% analytical uncertainty.

The analytical uncertainties in expressions (14) - (16) are presumably due to the uncertainty of G and $\omega_o$. To support this, let us insert $T_{max}$ = 14.7655 days into (16), obtaining for the certainty range:

$$T_o\big|_{Earth} \in [3419\,s,\ 3455\,s]. \quad (17)$$

Here the intensity of the interval (17), of $|T_o|$ = 18 s, reflects only the higher-order terms in Eq. (7), as well as the 17 s accuracy of the lunar synodic periodicity observation, and hence of the 8 s filtering step used, as well. Note that the above intensity falls within the certainty limits of SG-observed $\omega_o$'.

Of all the possible values from the above interval (17), only the upper-limit value is of interest here, as it is precisely this value that corresponds to the oscillation of the Earth's total mass. So:

$$T_o\big|_{Earth} = 3455\,s. \quad (18)$$

Inserting (18) back into the relationship (13) gives $2 \cdot 10^{-4}$ for the uncertainty of the relationship (13), clearly reflecting only the higher-order terms in Eq. (7), and the NIST relative standard uncertainty for $G_{Newton}$, of $1.5 \cdot 10^{-4}$. Thus the relationship (13) is verified within reason.

It is now justified to write the full relationship using (18) for the Newtonian scale, as:

$$\frac{\omega_o}{\omega_{max}} = C(\tau; T_{t.d.t.}), \quad C \in \Re, \quad (19)$$
$$C\big|_{Earth-Moon} = 369.2443415$$

$$\left.\begin{array}{l} G_{Newton} \cdot c \cdot C^I \to e^2 \pm 0.006\% \\ \\ G_{Planck} \cdot 10^{28} \cdot c \cdot C^{II} \to e^2 \\ \\ \dfrac{G_{Newton}}{G_{Planck}} \cdot \dfrac{1}{C^{III}} \to c^3, \end{array}\right\} \quad (20)$$

where the Roman superscript in the shorthand $C$ indicates the type of the oscillator of interest, with its own $\tau$-temporally varying resonance. Specifically:

$$\left.\begin{array}{l} C^I = \dfrac{\omega_o}{\omega_{max}}\bigg|_{Earth-Moon} \\ \\ C^{II} = C^I \pm \varepsilon_2; \quad \varepsilon_2 \in \Re \,\wedge \\ \quad \varepsilon_2 \ll \varepsilon_1 \\ C^{III} = C^{II} \pm \varepsilon_3; \quad \varepsilon_3 \in \Re \,\wedge \\ \quad \varepsilon_3 \equiv f(\varepsilon_1, \varepsilon_2; \tau) = \\ \quad \dfrac{\varepsilon_1}{\varepsilon_2} + \ldots \wedge \\ \quad \varepsilon_3 \gg \varepsilon_1. \end{array}\right\} \quad (21)$$

Note the excellent agreement for the Newtonian scales – first of the expressions (20). In the above, the notion of scales is induced such that the uncertainty $\varepsilon$ with which the proposed model fits the natural relationship on a given type of scales – represents that type of scales.

## 8. Discussion of the Corollary

If the Universe is a *hyperclock* of masterfully (i.e., with extremely slow decay) tuned X number of hyperclocks (from a string to the Universe), where X is an excruciatingly excruciatingly excruciatingly (…) large number but not infinity, then the above-obtained relation amongst the uncertainties

$$\varepsilon_3 \gg \varepsilon_1 \gg \varepsilon_2 \quad (22)$$

indicates that there is no real distinction between physics at the Newtonian ($\varepsilon_1$) and Planck ($\varepsilon_2$) scales. To examine the quality of the estimate of (18), let us substitute an equality $C^{III} = C^{II} = C^I$ into the remaining two of the expressions (20):

$$\left.\begin{array}{l} G_{Newton} \cdot c \cdot C^I \to e^2 \pm 0.006\% \\ \\ G_{Planck} \cdot 10^{28} \cdot c \cdot C^I \to e^2 \pm 0.252\% \\ \\ \dfrac{G_{Newton}}{G_{Planck}} \cdot \dfrac{1}{C^I} \to c^3 \pm 0.001\%. \end{array}\right\} \quad (23)$$

This made the analytical uncertainty in case of the Planck scales ($\varepsilon_2$) somewhat improve compared to (15) albeit not drastically. Thus, according to the proposed model, an inequality $C^{II} \neq C^I$ holds, meaning the first two types of scales bear different but similar properties. In case of the Einsteinean (intergalactic) scales ($\varepsilon_3$) however all of the analytical uncertainty practically vanished compared to (16), and equality $C^{III} = C^I$ holds. Hence, according to the model, physical properties at the Einsteinean scales do not differ from those at the Newtonian scales.

Then the intergalactic scales ($\varepsilon_3$) can be entirely described within the realms of the other two types of scales, while according to (20) scales smaller than $\varepsilon_2$ are not forbidden. This suggests that the true nature of all reality lays in the harmony of all oscillation, such as that proposed partially by string theory, and not in simplistic sums of infinitely many cases of "free fall", regardless of the curvature of "space-time" Riemannian representations of reality. (An analogy would be a straight line being just an infinitesimally small segment of a circle.) To arrive at this result a common principle at all scales — the resonant property (19) of gravitation — was used here.

The outcome of the proposed model is that there is no real distinction between the realities at different scales, while the proposed model still allows for scales smaller than Planck's to exist. Perceptions that are contrary to the proposed model however, such as the Einstein's "general relativity" with its claimed universality, appear to stem not from real distinctions but rather from remarkably different values of uncertainties in physical considerations carried out at the three classical types of scales.

Finally, the natural relationship for physics at the Newtonian and Planck scales reads, cf. (23):

$$\left. \begin{array}{l} G_{Newton} \cdot c \cdot C^I = e^2 \\ \\ G_{Planck} \cdot 10^{28} \cdot c \cdot C^{II} = e^2 \pm \varepsilon_2 \\ \\ \dfrac{G_{Newton}}{G_{Planck}} \cdot \dfrac{c^{-3}}{C^I} = 1 \, . \end{array} \right\} \quad (24)$$

## 9. Conclusion

In 1998 [26] stunned the scientific community by discovering that the Earth is able of vibrating at all times. To demonstrate this (for the high-frequency part of the natural band), [26] used recordings from the Japanese SG at Antarctica. Their discovery represented the first indication that [49] was correct when asserting that the Earth is a simple closed mechanical oscillator – in which case an external forcing could cause it to vibrate virtually incessantly, i.e., for as long as the external forcing lasts. In order to demonstrate this however the Earth must be shown as being able to vibrate incessantly in its low-frequency band, as well. This is what I have done at the outset.

Therefore a simple Earth-model is proposed based on the spectra of decadal gravity variations recorded by a superconducting gravimeter with 1Hz sampling rate. By testing for correlation between the mean-diurnal variance-spectra of SG-obtained Earth gravity, and the Earth's decadal strong (> M6.3) seismicity, a significant 3-day delay in such a correlation is identified. Mechanical-oscillator equations for the lunar synodic period (dynamically affecting all the Earth masses), the Earth mantle's usual viscosity, and the normal mode of Earth oscillation, successfully model an identical 3-day (±1%) phase delay. Further inspection into such a simple mechanical model's accuracy revealed that the proposed model also successfully predicted the maximum particle displacement on Earth as ~10 m, which is commonly reported as the average fault disruption during gravest global earthquakes.

The proposed model of the Earth taken as a simple mechanical oscillator forced primarily by the lunar gravitational pull suggested that the Earth tectonics is a mere consequence of the mantle's springtide-induced extreme resonant dynamics. Since – in the simplified world of the proposed model – tectonics is fully describable, large earthquakes are a consequence of structural collapses occurring when the magnified springtide-induced resonance matches the grave mode of oscillation of separate (tectonic) mass bodies in the upper mantle and in the crust. If the real Earth can be shown (for example: through a trial-and-error approach; or by computing the World Geoid where $C$ in (19) on the Newtonian scales is a function – first of Eqs.(24)) to satisfy the proposed model to a reasonably good approximation, then earthquake forecasting on real Earth would become a straightforward task.

Besides accounting for large discrepancies in the absolute values of *G* obtained in the past uncritically from (terrestrial) experiments at different locations and epochs, the springtide-induced magnification of the resonance of the Earth's mantle is likely responsible also for several still unexplained phenomena such as the well-established differential rotation of the Earth's inner core, the unexplained periodic portion of the Earth's nutation, as well as the Earth's dynamo i.e. magnetic field.

In order to show that the above model holds everywhere on Earth and at all times, the proposed model must be applicable to a realm that is beyond that of the Earth herself. And all that one has to do in order to ultimately demonstrate that the so-called reality (*undefined totality*) is nothing but a set of entangled series of series of series of (...) of mechanical oscillators, is to find a unique unifying relationship that relates physics at all scales.

Thus the *universal hyperclock* concept was herein examined. This was done in the only form such a hypothetical concept could reveal itself to us on Earth as absolutely measurable – that of the global mechanical oscillator model. This model was substantiated by superconducting gravimeter as the humankind's most accurate instrument. Note that this concept requires that permanent decay of all energy (permanent oscillation dampening) be at play, where no constants or units are generally permissible. Nor is the Einstein's equating the time and the clocks (i.e., atomic orbital periods) allowed, also. In the same sense, so-called "general relativity" merely represents another way of describing the dynamics of the $\varepsilon_1$ and $\varepsilon_2$ realities bounded by three real dimensions. (Note that (22) says nothing of scales smaller than $\varepsilon_2$, and of to them alleged higher dimensions, e.g., by string theory.)

Generally, imperfect relativitistic explanations such as the attempt to model the perihelion advance of the Mercury must not serve the purpose of proving illogical concepts such as the Einstein's so-called general relativity. On the contrary, observational facts must be used to construct viable theories [62]. According to the herein proposed model, properly modeling any gravitational-orbital phenomena (including perihelion advance, nutation), as observed for any object in a solar system, requires first the knowledge (measurement) of the grave mode of oscillation of all the total-masses which comprise the gravitational fields of both the object of interest and the with it oscillating (about it orbiting) objects. In this way it would be possible to examine with a certainty whether gravitational and electromagnetic fields could be equated using the mechanical-electrical resonance with resonant magnification of mass oscillation as a universal property of gravitation. To prove (20) would require setting up absolute and superconducting gravimeters on the surface of the objects of interest – the sun and all of about it orbiting bodies in case of our solar system.

If the herein proposed model could be verified in the above and other ways, and hence applied onto the real Earth and our solar system, the totality of all the masses in the Universe and beyond it – could be guessed. For this, resonant magnification of gravitation would need to be properly accounted for in all permutations of all existing gravitational (tidally locked) systems, regardless of the scale. Such a proof could deem redundant the elusive concepts of "dark energy" and "dark matter", as it would account for the totality of the entire Hyperverse and not only of its radio-observable part known as the Universe. By extension, the radio-observed mass acceleration of the Universe would then indicate that our Universe itself is entangled into the Hyperverse of oscillating universes, and that the whole of *matter attraction* is at the same time the *matter vibration* that at times is being excited and at times dampened. No so-called physical constant is absolutely constant then, but only apparently so – on the given scale and epoch.

In the proposed model the following relationship then generally holds, c.f. (24):

$$G = s \cdot e^2 \qquad (25)$$

or, writing the scaling factor, *s*, fully:

$$G_{Newton} = C^{-1} \cdot c^{-1} \cdot e^2$$
$$G_{Planck} = C^{-2} \cdot c^{-4} \cdot e^2, \quad (26)$$

which for the Earth-Moon system, (18) and (7), is well satisfied to within the NIST standard uncertainty for *G*. The Planck constant *h* is determined analogously to the principle of SG [63]. Thus, besides $G_{Newton}$, *h* too reflects the effect of the Earth total-mass (mostly the mantle) resonance, *C*, as given by Eq. (19). Hence the power of two in $C^{-2}$, Eq. (26), corroborates the proposed model. Note that, according to general understanding, the theory of quantum gravity allows for constants other than *c*, *G*, and *h*, such as *C*.

If *C* is constant then it is the only constant, in which case its value (19) is approximate. Then not only the Earth but also every spatially unique segment of the totality (*Hyperverse*) would maintain this constant, as well. The ratio of that segment's (say, the outer Universe's) grave mode of oscillation, and the orbiting period of another segment (outer outer Universe) from the Hyperverse orbiting about the outer Universe, would be a constant approximately equal to *C* found in (19) for the Earth. In case of the Earth, it is the Moon that largely masks the tidal influences of the rest of the Universe onto the Earth. In case of our Universe, it is to it neighboring (about it orbiting) Universe that largely masks the tidal influences of the rest of the Hyperverse onto our Universe...

The above-proposed space mission to verify (20) would show whether *C* and *c* in (26) are both universal constants, or, on the contrary, *G* (and hence physics) changes with location. In either case the (discrete or continuous) generalised form of (26) gives

$$G_{string} = C^{-3} \cdot c^{-7} \cdot e^2 = 6.74377 \cdot 10^{-67} \quad (27)$$

Finally, the most general form of (26) reads:

$$G_{Hyperverse} = c^2 \cdot e^2 = 6.64095 \cdot 10^{17}, \quad (28)$$

where the 0-th power of *C* means harmony of all oscillation. Non-fractional *C* are obviously forbidden in local real domains. Therefore, further generalisations of (26) are not possible for scales larger than that of (28).

Within the above-asserted (Hyperverse) extension of the proposed global oscillator model onto everything everywhere, the reality could be represented by a lattice of strings (hyperverses to us) in an even larger-scale *Oberverse*. By extension, there is nothing in the proposed model that would forbid the lattice of to-us strings from being a hyperverse of even smaller-scales' worlds, and so on. The logic of infinity seems inconceivable to us as the infinity could correspond not to our (linear) perception of the frequency domain but to that entire domain itself, instead.

Three linear dimensions bind our consciousness. As a result, we are incapable of relating ourselves to the reality in terms of frequency space which then is forever non-obvious (hidden) to our senses, but not to our instruments or analysis methodologies such as spectrography. Thus, gravity is not a force as Newton linearly misinterpreted the frequency-space reality; nor is gravity geometry as Einstein linearly misinterpreted the same reality – arguably, less inaccurately than Newton did. Instead, gravity is omnipresent, and it does not "act on distance" by some illusive "gravity waves". Rather, *gravitation is vibration*.

The classical (Newtonian-Einsteinian) view is that gravitation can be only attractive, as based on our everyday geocentric experience. However, this is only apparently so. Rather, we must get rid of the geocentric view altogether. Then, looking from all the points in the outer space simultaneously in the direction of the Earth, since gravitation is vibration its influence spreads forward (repulsively) throughout the space and in all directions, from all the points that have their own mass/energy manifestation, e.g., particle spin. For this, no special particles so-called "gravitons" are needed; instead, gravitation is a never-ending influx of the gravitational waves that are disturbing (i.e., vibrating along) the 3D aether so-called "dark matter (energy)". Thus, what is usually referred to as the "Earth gravitation", in fact is the resultant of the aether-disturbing waves arriving to the Earth from the whole of the Universe onto the observer at a point. This means that all mass/energy of points (i.e., orbital energy of particles) release parts of their momenta, causing a disturbance in the aether resulting in waveform deformations of aether's steady states. <u>This disturbance takes form of the (gravitation's) mechanical waves</u>. An evident modulation of this disturbance occurs during the full and new Moon, when the Moon obscures the normal (line-of-sight-) direction of the aether disturbances as those disturbances are incoming from whole of the Hyperverse (the Moon and the Sun being the largest concentration of particles nearby). This instantaneous obstruction makes then the vibrational nature of gravitation obvious on Earth during full and new moon, as well as during the eclipses.

  In case of the Earth for instance, the totality of subatomic orbits in all the particles composing the Earth transfers a significant amount of their orbital momenta onto Earth's surroundings, due to the subatomic particles acting as forced mechanical oscillators just like the Earth-Moon system does. The totality of transfer of momenta from all subatomic particles' orbits onto their domicile body of mass (local group of particles) determines that body's grave mode of oscillation, here called $\omega_{\text{body own resonance}}$. The higher the concentration of particles at a spatial locality means the more massive (energetic) the observed concentration of particles, such as the Earth body. It also means a greater push exerted by all of the Hyperverse points onto the observed body, i.e., the stronger the apparent so-called "gravitational field" of the body. Hence, the so-called central gravity fields in the Newtonian theory are only apparent. Thus, the ratio

$$\frac{\omega_{\text{Earth-Moon}}}{\omega_{\text{Earth}}} \quad (29)$$

approximately determines at all times the value of the scaling parameter of physics, G, at the appropriate scale. This is written in general form as

$$\frac{\omega_{\text{body orbital resonance}}}{\omega_{\text{body own resonance}}}. \quad (30)$$

When a heavenly body has no about it orbiting objects, then the period in the numerator takes value $T = 2\pi$, and the gravitation-determining ratio takes form

$$\frac{1}{\omega_{\text{body own resonance}}}. \quad (31)$$

When a heavenly body, say the Sun has $N$ about it orbiting objects (the so-called $N$-body problem from classical mechanics), then the gravitation-determining ratio is

$$\frac{N \cdot \omega_{\text{body orbital resonance}}}{\left(\omega_{\text{body own resonance}}\right)^N}. \quad (32)$$

The instant when this ratio in all local particles is about to approach the constant C, the process known as the *accretion* begins, and a new heavenly body starts to form in the observed locality. The physical meaning of the radial component of gravitation, as established in the above, is as follows.

  It is established in the above that gravitation propagates via mechanical waves in the medium (aether) only outward; hence the famous 'minus' sign in the Newton's law is obsolete. The radiating propagation of gravitation gives also rise to what is generally observed as the expansion of the Universe. The aether is composed of charged matter, mislabelled as "dark matter" or "dark energy" in an attempt to account for a 98%-misfit between astrophysical observations and the current working model of physics based on the Einstein's so-called General Relativity theory. The Earth lunar tides are the most obvious manifestation of the repulsive propagation of gravitation's mechanical waves throughout the aether. Namely, the high and low tides, appearing respectively on the sides of the Earth directly facing and directly opposite from the Moon, are caused by the Moon obstructing the balance of overwhelming totality of gravitational waves that emanate from the whole of Hyperverse (thus normally pressing onto the Earth to form the Earth's spherical shape). A tide is not a "pull exerted by one body of mass onto another body of mass"; this is a magical rather than a scientific explanation, and I believe it stems entirely from Newton's obsession with mysticism.

Instead, a tide is an instantaneous imbalance of the spherical shape of a body, i.e., of the totality of gravitational waves arriving from the entire Hyperverse to a body therefore normally balanced to form a sphere. The tidal distortion occurs when the body's near-orbiting satellite obstructs the locally apparent line-of-sight between the body and the gravitation's mechanical waves arriving from beyond the body's satellite (that is: from a cone with the tip at the location of the observer on Earth, and open to deep space behind the satellite).

### SUMMARY OF MOST IMPORTANT PROPORTIONS

$$\left| G_{Newton} \right|_{NIST} = (6.6742 \pm 0.001) \cdot 10^{-11}$$

$$\left| G_{Planck} \right|_{NIST} = (6.7087 \pm 0.001) \cdot 10^{-39}$$

$w_{o\ Earth} / w_{Earth-Moon}$ = 14.7655 day / 3455 sec = $C$ = 369.2443415

$$\left| c^{-1} \right| \cdot C^{-1} \cdot e^2 = 6.6750 \cdot 10^{-11} = \left| G^*_{Newton} \right|$$

$$\left| c^{-4} \right| \cdot C^{-2} \cdot e^2 = 6.7093 \cdot 10^{-39} = \left| G^*_{Planck} \right|$$

$$G = s \cdot e^2$$

$$G_{string} = C^{-3} \cdot c^{-7} \cdot e^2 = 6.74377 \cdot 10^{-67}$$

$$G_{Hiperverse} = C^0 \cdot c^2 \cdot e^2 = 6.64095 \cdot 10^{17}$$

…

---

Or, written differently:

$$G^*_{Hyperv.} = (\ )^0 e^2 c^2 = 6.64095 \cdot 10^{17}$$

$$G^*_{Newton} = \left( \frac{\omega_{Earth-Moon}}{\omega_{Earth}} \right)^1 \cdot \frac{e^2}{c} = 6.6750 \cdot 10^{-11} = G^{NIST}_{Newton}$$

$$G^*_{Planck} = \left( \frac{\omega_{Earth-Moon}}{\omega_{Earth}} \right)^2 \cdot \frac{e^2}{c^4} = 6.7093 \cdot 10^{-39} = G^{NIST}_{Planck}$$

$$G^*_{strings} = \left( \frac{\omega_{Earth-Moon}}{\omega_{Earth}} \right)^3 \cdot \frac{e^2}{c^7} = 6.74377 \cdot 10^{-67}$$

…

arXiv.org44. Merriam, J. Atmospheric pressure and gravity. *Geoph. J. Int.* **109** (3), 488-500 (1992).
45. Smylie, D.E., Hinderer, J., Richter, B. & Ducarme, B. The product spectra of gravity and barometric-pressure in Europe. *Phys. Earth Planet. Interiors* **80** (3-4), 135-157 (1993).
46. Tamura, Y. A harmonic development of the tide-generating potential. *Bull. Inform. Mar. Terr.* **99**, 6813–6855 (1987).
47. Chapront-Touzé, M. & Chapront, J. *Lunar Solution ELP 2000-82B*. Willmann-Bell, Richmond, Virginia (1988).
48. Hendershott, M.C. Oceans: waves of the sea. In: Encyclopædia Britannica CD (1994).
49. Tesla, N. The magnifying transmitter. *The Wireless Experimenter*, June (1919).
50. Den Hartog, J.P. *Mechanical Vibrations*. Dover Publications, New York (1985).
51. King, S. D. Models of mantle viscosity. In: Ahrens, T. J. (Editor). AGU Handbook of Physical Constants (1994).
52. Stein, R.S. Tidal triggering caught in the act. *Science* **305**, 1248-1249 (2004).
53. Cisternas, M., Atwater, B.F., Torrejón, F., Sawai, Y., Machuca, G., Lagos, M., Eipert, A., Youlton, C., Salgado, I., Kamataki, T., Shishikura, M., Rajendran, C.P., Malik, J.K., Rizal, Y. & Husni, M. Predecessors of the giant 1960 Chile earthquake. *Nature* **437**, 404-407 (2005).
54. Felzer, K. R. & Brodsky, E. E. Decay of aftershock density with distance indicates triggering by dynamic stress. *Nature* **441**, 735-738 (2006).
55. Rubin, K.H., Van der Zander, I., Smith, M.C. & Bergmanis, E.C. Minimum speed limit for ocean ridge magmatism from 210Pb-226Ra-230Th disequilibria. *Nature* **437**, 534-538 (2005).
56. Sieh, K., Stebbins, C., Natawidjaja, D.H. & Suwargadi, B.W. Mitigating the effects of large subduction-zone earthquakes in Western Sumatra. Abstract PA23A-1444, AGU Fall Meeting, December 2004, San Francisco (2005).
57. Thatcher, W. Seismic triggering and earthquake prediction. *Nature* **299**, 12 - 13 (1982).
58. Hanson, K., Maul, G.A. & Mcleish, W. Precipitation and the lunar synodic cycle phase progression across the United States. *Climate and Applied Meteorology* **26**, 1358-1366 (1989).
59. Lethbridge, M.D. Thunderstorms, cosmic rays, and solar-lunar influences. *J of Geoph. Res.* **95**, 13645-13651 (1990).
60. Benioff, H. Long waves observed in the Kamchatka earthquake of November 4, 1952. *J. Geoph. Res.* **63**, 589-593 (1958).
61. McCarthy, D.D. & Petit, G. (Editors). International Earth Rotation and Reference Systems Service. IERS Conventions 2003, Technical Note 32 (2003).
62. Wrinch, D. & Jeffreys, H. The relation between geometry and Einstein's theory of gravitation. *Nature* **106**, 806 – 809 (1921).
63. Williams, E.R., Steiner, R.L., Newell, D.B. & Olsen, P.T. Accurate measurement of the Planck constant. *Phys. Rev. Lett.* **81**, 2404–2407 (1998).
64. Van Camp, M. Measuring seismic normal modes with the GWR C021 superconducting gravimeter. *Phys. Earth Planet. Interiors* **116** (1-4), 81-92 (1999).
20

APPENDIX

**Supporting information**

In order to verify the grave mode of the Earth's total-mass oscillation, I take the SG gravity recordings during three greatest earthquakes from the 1990-ies that affected the Cantley SG record, Table A. Using the to GVSA unique ability to process gapped records, I look for the differences between spectra of the selected gravity recordings without gaps and with gaps artificially introduced in those records. I thus make 5, then 21, and 53 filter-step-long (8-32 sec) gaps in the three records respectively, where the order of earthquakes was selected at random. By observing the differences between the Gauss-Vaníček (G-V) spectra of complete *vs.* incomplete records, I look for the first instance when this difference reaches the zero value. Since both complete and incomplete records always described the same instance and same location when and where the same field (in this case the Earth gravity field) was sampled during the three energy emissions, it is precisely this value that marks the beginning of the Earth's natural band of oscillation. If this setup is correct, then the more gaps in the record should mean the more pronounced impact of the non-natural information onto the spectra. Indeed, Figure A shows that more gaps results in clearer distinction between the natural and non-natural band.

The effect of any known (including tidal) variation can be suppressed in the GVSA together with processing, within *enforcing*, see [17], where known periods in form of analytical functions or discrete data sets can be forced on data. Thus, when GVSA is used, no preprocessing is required for the purpose of stripping the observations of tides. Here nine semidiurnal and diurnal tidal periods were enforced[*]: 13.4098257, 13.9426854, 14.5057965, 15.0424341, 15.5742444, 28.4395041, 28.9840259, 29.5159428, and 29.9977268 º/hr. Spectral lines superimposed on the spectral plots on Figure A represent normal mode periods from the Zharkov's model [41].

Thus the grave mode (most natural period) of the Earth total mass oscillation is measured as $T_o' = 3445$ s ±0.35%, where uncertainty is based on 1000 pt spectral resolution. This stands in staggering agreement with [60], although the seismological community has been very critical of that estimate, mulishly insisting on "noise" removal.

| Earthquake event order | date | $M_S$ | Approximate location | φ | λ | $d$ km |
|---|---|---|---|---|---|---|
| #48 | 11/08/97 | 7.9 | China | 35.07 | 87.32 | 33 |
| #79 | 03/25/98 | 8.8 | Ballenys, South of Australia | -62.88 | 149.53 | 10 |
| #80 | 11/29/98 | 8.3 | Pacific Ocean, Indonesia | 2.07 | 124.89 | 33 |

Table A. Three strongest shallow earthquakes in the Cantley record from 1990-ies.

---

[*] Hou, T. *Sequential tidal analysis and prediction*. MSc Thesis, Department of Geodesy and Geomatics Engineering, University of New Brunswick, Fredericton, Canada (1991).

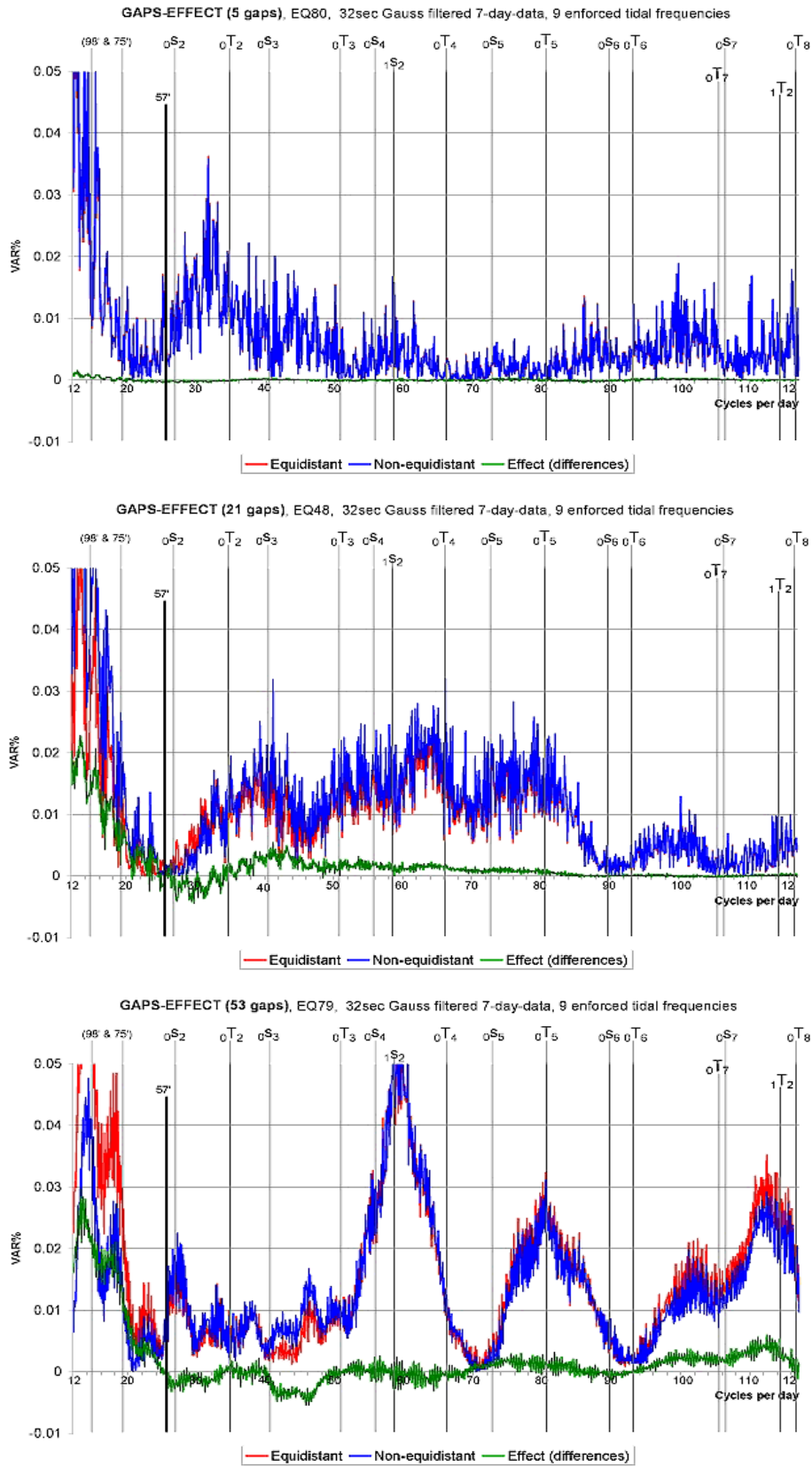

Figure A: The effect (green) of regarding a series as equidistant, for 5, 21 and 53, 8-sec gaps. Shown are G-V spectra of gravity at one week past 3 strongest shallow earthquakes, Table A.